\documentclass[
  aps,
  pre,
  reprint,
  floatfix
]{revtex4-2}

\usepackage{amsmath,amssymb,amsfonts}
\usepackage{graphicx}
\usepackage{booktabs}
\usepackage{float}
\usepackage{tikz}
\usepackage{color}
\usepackage{siunitx}
\usepackage[percent]{overpic}
\usepackage{hyperref}

\begin{document}

% -----------------------------
% Title
% -----------------------------
\title{Determining critical temperature
differences of low-temperature-differential Stirling engines:
nonlinear dynamics approach }

% -----------------------------
% Authors
% -----------------------------
\author{Momoko Amemiya}
\thanks{1092054156@edu.k.u-tokyo.ac.jp}
\affiliation{Department of Complexity Science and Engineering, Graduate School of Frontier Sciences, The University of Tokyo, Kashiwa 277-8561, Japan}

 \author{Yuki Izumida}
 \thanks{izumida@k.u-tokyo.ac.jp}
 \affiliation{Department of Complexity Science and Engineering, Graduate School of Frontier Sciences, The University of Tokyo, Kashiwa 277-8561, Japan}

% -----------------------------
% Abstract
% -----------------------------
\begin{abstract}
While the low-temperature-differential (LTD) Stirling engines are innovative engines that can operate with low temperature differentials in our daily life, the problem of determining the critical temperature differences below which the engine ceases to rotate remains unexplored.
In this study, we solve this problem using a nonlinear dynamics approach.
We derive the self-consistent equations that determine the critical temperature differences as homoclinic bifurcation points of a dynamical model of the LTD Stirling engines.
The solutions of the self-consistent equations reveal a combination of parameters that determines the critical temperature differences.
This enables us to establish the fundamental design principles for improving the performance of the LTD Stirling engines beyond empirical designs.
\end{abstract}
\maketitle

% -----------------------------
% Main text
% -----------------------------
\section{Introduction}
Thermal energy technology is becoming increasingly important as we face growing energy demand by population growth and the rapid development of AI~\cite{Chen2025}.
The low temperature heat is one of the key energy resources as it is ubiquitous in our daily life or it is available as waste heat from plants and data centers, which is not fully utilized due to technological difficulty.

The low-temperature-differential (LTD) Stirling engine enables the utilization of such low temperature heat, which was invented by Ivo Kolin in the 1980s~\cite{Senft2000,Senft2010}.
It was a technological innovation in the history of Stirling engines.
A model with sophisticated design can operate with a temperature difference as small as that between body and room temperatures, demonstrating its potential use for a sustainable society.
Despite its importance, the LTD Stirling engine has yet to be in practical use mainly due to the low power output.

The LTD Stirling engine has a critical temperature difference below which its motion ceases.
The critical temperature difference varies depending on the design, and decreasing its magnitude contributes to improve the performance of the LTD Stirling engines.
Since its invention, the engine's design was improved and refined in order to decrease the critical temperature difference~\cite{Senft2000,Senft_2007,Senft_1993}.
Nevertheless, these improvements were largely based on experiences and it remains unclear which combinations of physical quantities determine it. 

In this paper, we determine the critical temperature differences of the LTD Stirling engine based on a nonlinear dynamics approach.
Along with experimental demonstrations~\cite{KONGTRAGOOL2007547,lesack2025realtimequantitativemeasurementstirling,10.1119/1.5141965,toyabe2020experimental,Ruijie},
theoretical dynamical modeling of the engine is a powerful tool for describing the engine's dynamics and its rotational mechanism~\cite{Robson2007,Craun2017,izumida2018nonlinear,Romanelli}. The simplest dynamical modeling with a few degrees of freedom~\cite{izumida2018nonlinear} revealed a bifurcation mechanism behind the critical temperature difference, which was verified experimentally~\cite{toyabe2020experimental}.
We derive a self-consistent equation that determines the bifurcation point by using a standard technique of nonlinear dynamics. From the analysis of the solution of the equation, we elucidate the principal physical parameters that determine the critical temperature differences. This establishes a novel design principle for the LTD Stirling engine without relying on the ingenuity of inventors.

\section{LTD Stirling engine}
\begin{figure}[t!]
  \centering
  \includegraphics[width=0.9\linewidth]{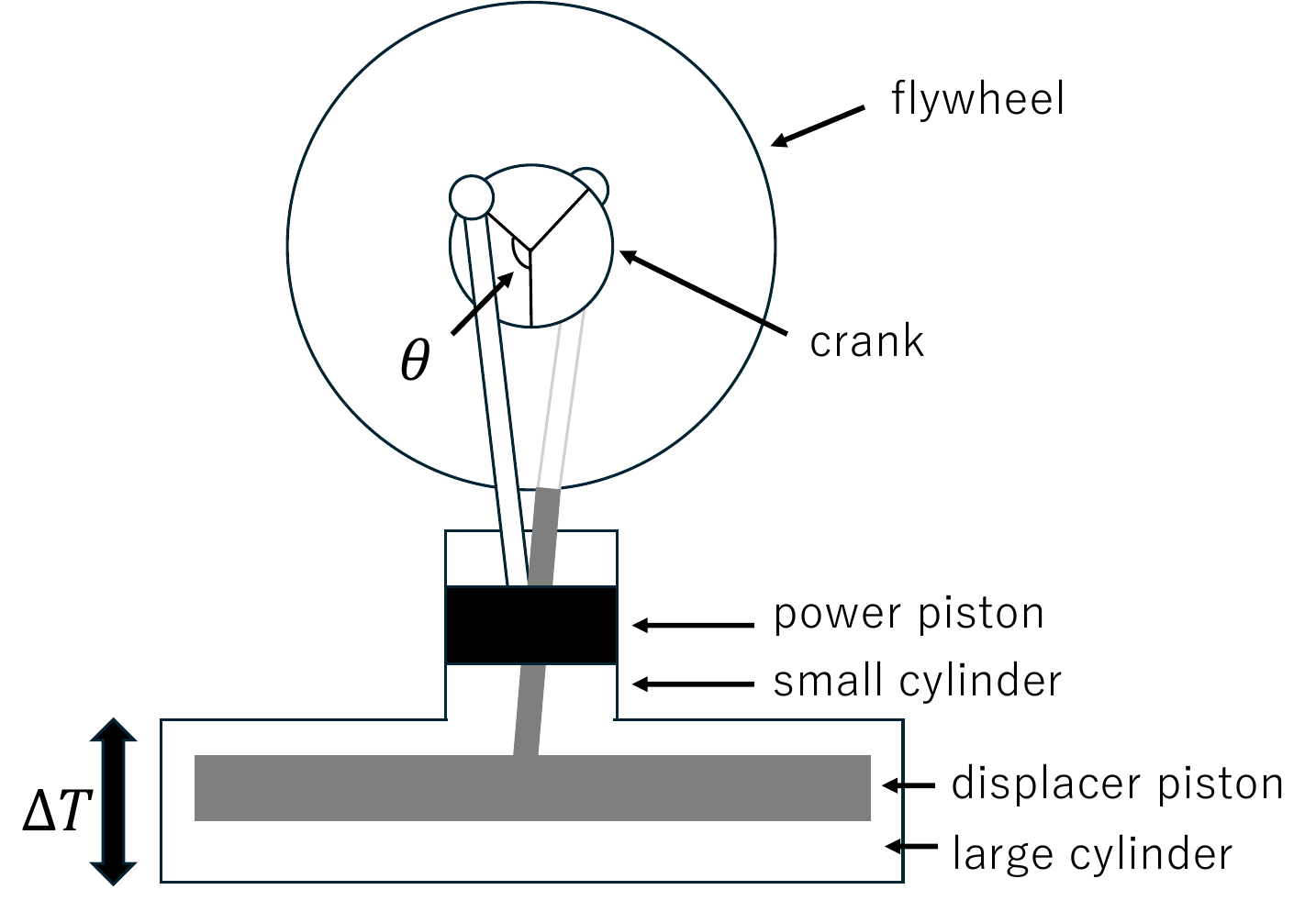}
  \caption{Schematic illustration of an LTD Stirling engine. The engine rotates via a piston-crank mechanism with the temperature difference $\Delta T$ (see the main text).}
  \label{fig:engine_picture}
\end{figure}

Figure~\ref{fig:engine_picture} shows a schematic diagram of an LTD Stirling engine, which has a gamma-type configuration~\cite{KONGTRAGOOL2003131}.
The LTD Stirling engine utilizes a piston-crank mechanism, which converts the reciprocating motion of the  piston into the rotational motion of the crank.
The working gas fills the cylinder that consists of the connected large and small cylinders. The large cylinder is in contact with heat reservoirs at its top and bottom, whose temperatures are denoted by $T_{\rm top}$ and $T_{\rm bottom}$, respectively.
Both the power piston and the displacer piston are connected to the crank, and the angle of the crank connected to the power piston, denoted by $\theta$, lags behind that connected to the displacer piston by $\pi/2$.
The displacer piston brings the working gas into contact alternately with the heat reservoirs at $T_{\rm top}$ and $T_{\rm bottom}$.
The power piston serves as the motive power of the engine.

When the magnitude of the temperature difference $\Delta T \equiv T_{\rm bottom}-T_{\rm top}$ is sufficiently large, the LTD Stirling engine rotates autonomously. 
The direction of the rotation reverses depending on
the sign of $\Delta T$.
When $\Delta T>0$, the engine rotates clockwise;
as the displacer piston moves upward, the working gas moves to the bottom side of the large cylinder, where it is heated, increasing its pressure.
The power piston is pushed up by the gas pressure, and the flywheel rotates clockwise via the crank. 
Accompanied by the flywheel rotation, the displacer piston moves downward. The working gas moves to the top side of the large cylinder, where it is cooled, decreasing its pressure. The power piston is pushed down against the gas pressure and the flywheel continues to rotate.
When $\Delta T<0$, the engine rotates counterclockwise in the opposite manner.

\section{Model}
We use the LTD Stirling engine model proposed in~\cite{izumida2018nonlinear, toyabe2020experimental}.
The model describes the engine's rotational motion using the equation of motion for a pendulum driven by the temperature difference.
The theory describes the crank angle and the angular velocity of the engine $(\theta,\omega)$:
\begin{align}
&\dot{\theta} = \omega, 
\label{eq.dim_model_theta}\\
&I\dot{\omega} =
\sigma_p[p(\theta,\omega)-p_{\rm air}]r\sin\theta-\Gamma\omega.\label{eq.dim_model_omega}
\end{align}
Here, $T(\theta, \omega)$, $p(\theta, \omega)$, and $V(\theta)$ denote the gas temperature, the gas pressure, and the volume of the connected large and small cylinders:
\begin{align}
&T(\theta,\omega) \equiv T_{\rm top}+\frac{1+\alpha\sin(\theta-\omega\tau)}{2}\Delta T, \\
&p(\theta,\omega) \equiv \beta\frac{nRT(\theta,\omega)}{V(\theta)},\\
&V(\theta)\equiv 2r\sigma_d+r\sigma_p(1-\cos\theta).
\end{align}
Here, $I$ and $\Gamma$ are the moment of inertia and the friction coefficient.
$r$, $\sigma_d$, and $\sigma_p$ are the crank radius, the sectional area of the large cylinder, and the sectional area of the power piston.
One of the characteristics of the LTD Stirling engine is a small compression ratio~\cite{Senft2000}, 
which implies $\sigma_p \ll \sigma_d$. 
$p_{\rm air}$, $n$, and $R$ are the atmospheric pressure, the number of moles of the internal gas, and the gas constant. 
$\alpha$, $\beta$, and $\tau$ are empirical parameters determined by fitting with experimental data~\cite{toyabe2020experimental}.
$\alpha \le 1$ represents the fraction of the applied temperature difference that effectively acts on the working gas,
and $\tau \ge 0$ describes the effect of time delay.
The temperature and pressure may not be uniform in the cylinder, so the internal gas is not assumed to obey the ideal gas law and $\beta$ quantifies such an effect. 
The ideal case with $\alpha=1$, $\beta=1$, and $\tau=0$ reproduces the model in~\cite{izumida2018nonlinear}.

\begin{figure}[tbp]
\begin{center}
\begin{minipage}[c]{0.72\linewidth}
  \centering
  \begin{overpic}[width=\linewidth]
    {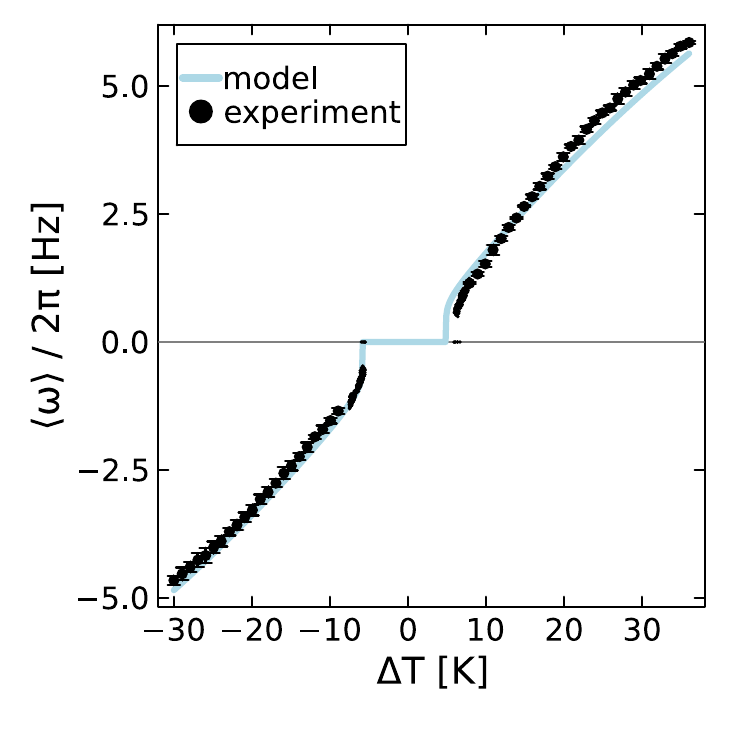}
  \end{overpic}
\end{minipage}

\vspace{2mm}
\begin{minipage}[c]{0.49\linewidth}
  \centering
  \begin{overpic}[
    width=\linewidth,
    trim=17 0 5 0,
    clip
  ]{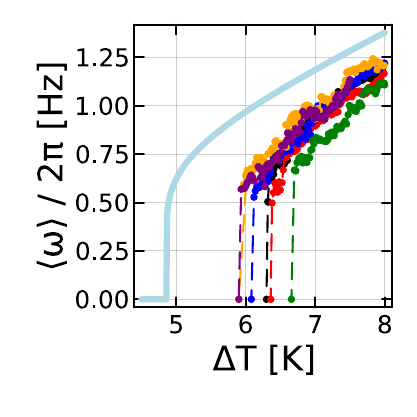}
  \end{overpic}
\end{minipage}
\hfill
\begin{minipage}[c]{0.49\linewidth}
  \centering
  \begin{overpic}[
    width=\linewidth,
    trim=17 0 5 0,
    clip
  ]{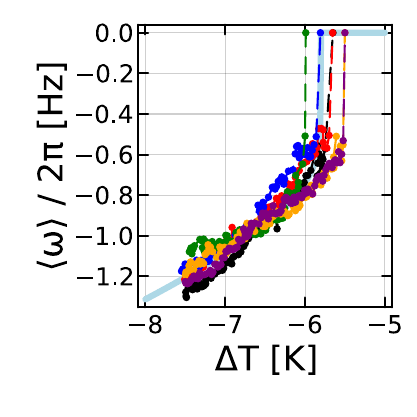}
  \end{overpic}
\end{minipage}

\end{center}

\caption{
Experimental data on an N-92 type LTD Stirling engine
(Kontax Engineering Ltd) and model-predicted values of
the time-averaged frequency $\langle\omega\rangle/(2\pi)$
at each temperature difference.
Data taken from Fig.~2(a) of Ref.~\cite{toyabe2020experimental}.
}
\label{fig:experiment_model}
\end{figure}

The rotational and stationary states of the engine correspond to a stable limit cycle and a stable fixed point of the dynamical system (\ref{eq.dim_model_theta}) and (\ref{eq.dim_model_omega}), respectively.
Figure~\ref{fig:experiment_model} shows a comparison of experimental data reported in~\cite{toyabe2020experimental} and model-predicted values of the time-averaged frequency as a function of $\Delta T$.
Here, we used the same physical parameter values as~\cite{toyabe2020experimental} and the empirical parameters $\alpha$, $\beta$, and $\tau$ are determined by maximum likelihood estimation.
The details of the numerical calculations are given in Appendix~\ref{app:plot_value}.

Since the LTD Stirling engine is driven by the temperature difference, decreasing its magnitude weakens the driving force. Below a critical temperature difference, the driving force can no longer overcome friction, and the engine ceases to rotate (Fig.~\ref{fig:experiment_model}).
This corresponds to the disappearance of the stable limit cycle of Eqs.~(\ref{eq.dim_model_theta}) and (\ref{eq.dim_model_omega}) via a homoclinic bifurcation.
A homoclinic bifurcation occurs when a part of a stable limit cycle collides with a saddle point and a homoclinic orbit is formed simultaneously~\cite{Strogatz2001}.
Here, the homoclinic orbit is defined as the orbit that recedes from the saddle point along the unstable manifold and approaches the same saddle point along the stable manifold after infinite time.

\begin{figure}[t]
  \centering

  \setlength{\fboxsep}{0pt}
  \setlength{\fboxrule}{0pt}

    \begin{minipage}[c][0.17\textheight][c]{0.46\linewidth}
      \centering
       \includegraphics[width=\linewidth]{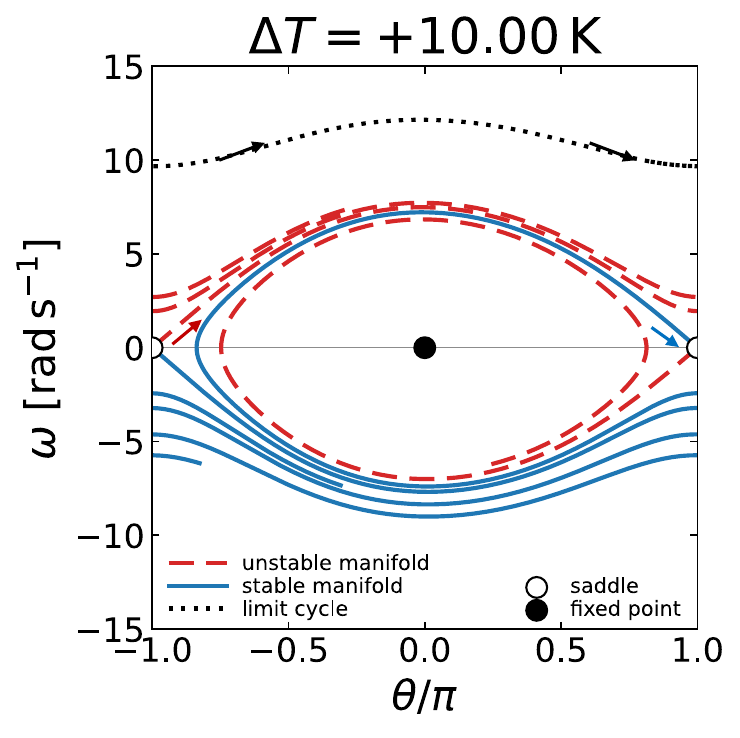}
    \end{minipage}
  \hfill
    \begin{minipage}[c][0.17\textheight][c]{0.46\linewidth}
      \centering
     \includegraphics[width=\linewidth]{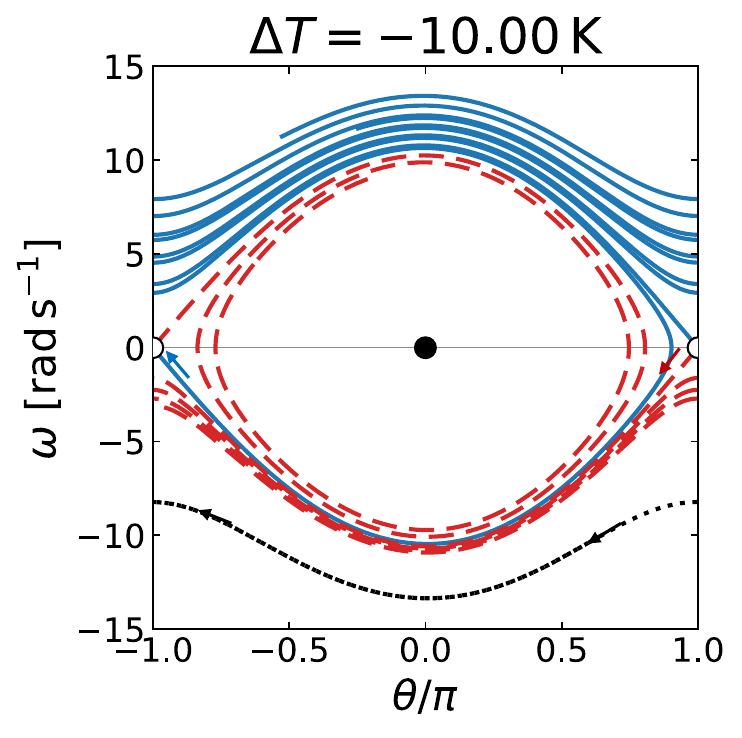}
    \end{minipage}

  \vspace{2mm}

    \begin{minipage}[c][0.17\textheight][c]{0.46\linewidth}
      \centering
         \includegraphics[width=\linewidth]{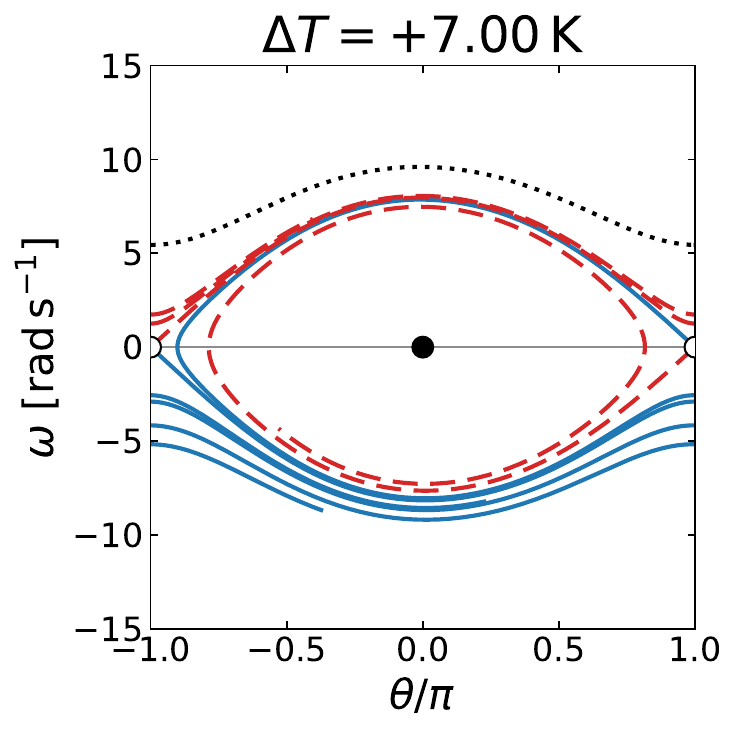}

    \end{minipage}
  \hfill
    \begin{minipage}[c][0.17\textheight][c]{0.46\linewidth}
      \centering
      \includegraphics[width=\linewidth]{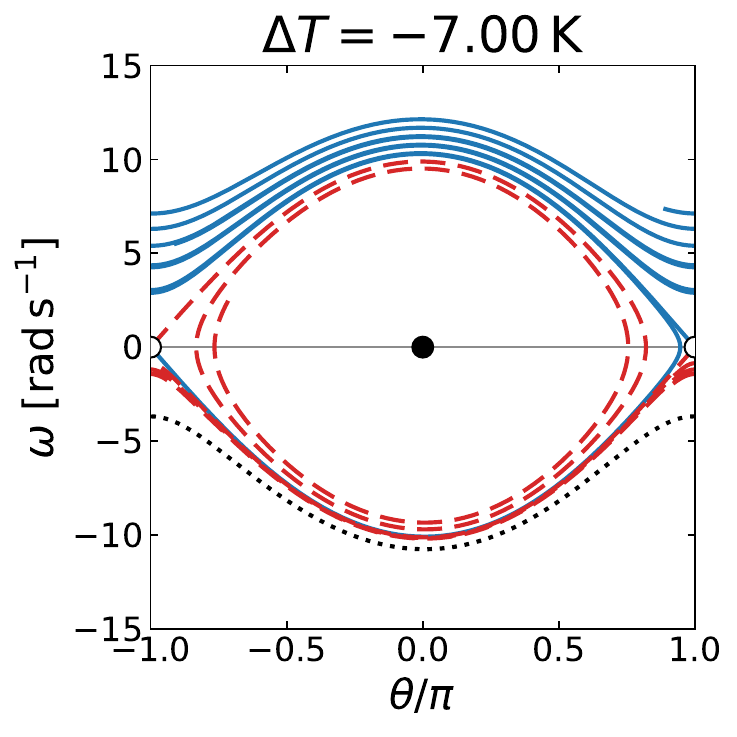}
    \end{minipage}

  \vspace{2mm}

    \begin{minipage}[c][0.17\textheight][c]{0.46\linewidth}
      \centering
        \includegraphics[width=\linewidth]{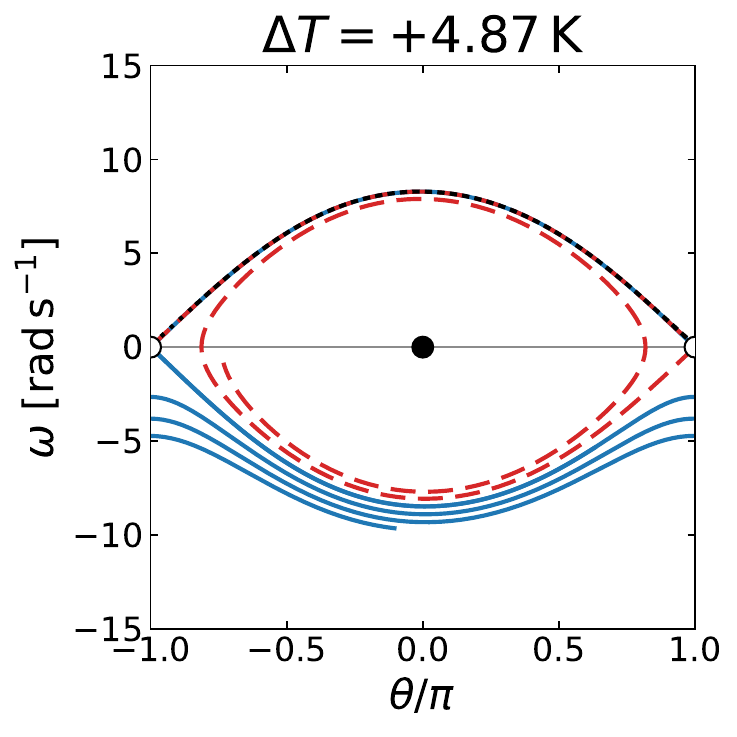}
    \end{minipage}
  \hfill
    \begin{minipage}[c][0.17\textheight][c]{0.46\linewidth}
      \centering
       \includegraphics[width=\linewidth]{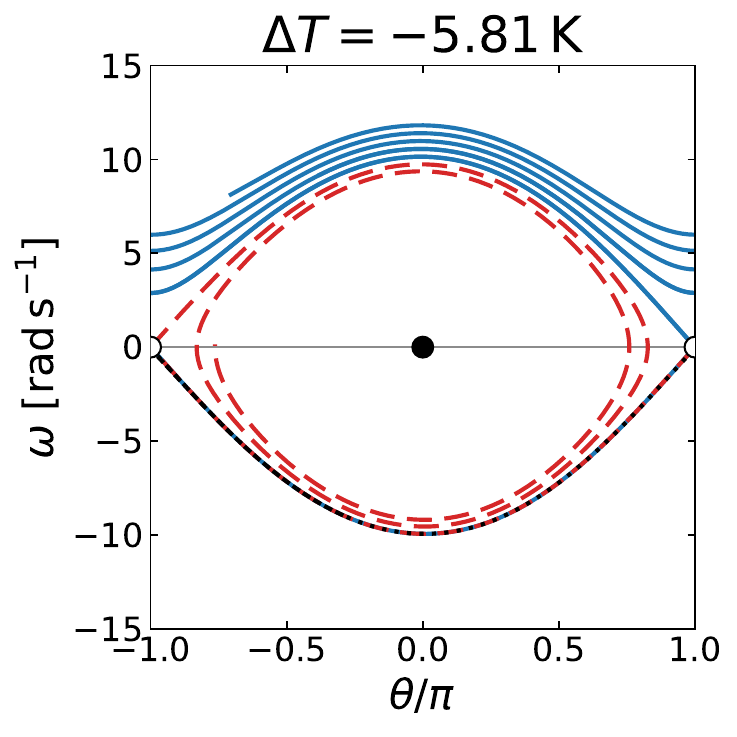}
    \end{minipage}

  \caption{Typical limit-cycle orbits in the vicinity of the critical temperature differences for (left) $\Delta T >0$ and (right) $\Delta T < 0$. For $\Delta T>0$, the critical temperature difference is $\Delta T_c^+ \simeq \SI{4.86}{K}$ and we plot $\Delta T=\SI{10}{K}$, $\SI{7}{K}$, and $\SI{4.87}{K}$.  For $\Delta T<0$, the critical temperature difference is $\Delta T_c^-\simeq\SI{-5.80}{K}$ and we plot $\Delta T=\SI{-10}{K}$, $\SI{-7}{K}$, and $\SI{-5.81}{K}$. 
  The black dotted curve represents the limit cycle.
  The open and filled circles represent the saddle point $(\pm \pi, 0)$ and the stable fixed point $(0, 0)$, respectively.
  The red dashed and blue solid curves represent the unstable and stable manifolds of the saddle point $(\pm \pi, 0)$, respectively. The small arrows indicate the direction of the time evolutions of the orbits.
  The same parameter values as Fig.~\ref{fig:experiment_model} are used (Appendix~\ref{app:plot_value}).
  }
  \label{fig:phase_diagram}
\end{figure}

Figure~\ref{fig:phase_diagram} shows the typical limit-cycle behaviors toward the critical temperature differences.
Here, $(\theta,\omega)=(\pm \pi,0)$ is the saddle point and $(\theta,\omega)=(0,0)$ is the stable fixed point of Eqs.~(\ref{eq.dim_model_theta}) and (\ref{eq.dim_model_omega}), where $(\pi,0)$ and $(-\pi,0)$ are identified as the same point due to $2\pi$-periodicity.
Figure~\ref{fig:phase_diagram} indicates that as the temperature difference approaches the critical value, a part of the stable limit cycle gradually approaches the saddle point $(\pm \pi, 0)$.
At the critical temperature difference, the unstable manifold of the saddle is connected to its stable manifold forming the homoclinic orbit, and the stable limit cycle disappears~\cite{Strogatz2001}.
We denote the critical temperature differences for $\Delta T>0$ and $\Delta T<0$ by $\Delta T_c^+$ and $\Delta T_c^-$, respectively (Fig.~\ref{fig:experiment_model}).

\section{Main results}
We used the Melnikov method to obtain the critical temperature differences of the LTD Stirling engine.
The Melnikov method is a standard technique to obtain an analytic expression for a homoclinic bifurcation point approximately~\cite{Guckenheimer1983,Wiggins2003}.
For convenience, we nondimensionalize Eqs.~(\ref{eq.dim_model_theta}) and (\ref{eq.dim_model_omega})
by introducing
$\tilde t\equiv\sqrt{\frac{nRT_{\rm top}}{I}}t,\tilde\tau\equiv\sqrt{\frac{nRT_{\rm top}}{I}}\tau,\tilde\omega\equiv\frac{\omega}{\sqrt{\frac{nRT_{\rm top}}{I}}},\tilde\sigma\equiv\frac{\sigma_p}{\sigma_d},\tilde\Gamma\equiv\frac{\Gamma}{\sqrt{nRT_{\rm top}I}},\tilde p_{\rm air}\equiv\frac{\sigma_drp_{\rm air}}{nRT_{\rm top}},\tilde T\equiv\frac{T}{T_{\rm top}}$ , and $\Delta\tilde T\equiv\frac{T_{\rm bottom}-T_{\rm top}}{T_{\rm top}}$, where quantities with tildes denote nondimensionalized ones.
The resulting nondimensionalized equations~(\ref{eq.dim_model_theta}) and (\ref{eq.dim_model_omega}) read:
\begin{align}
\dot{\theta} &= \tilde\omega, \label{eq.nodim_model_theta}\\
\dot{\tilde{\omega}}&=\tilde\sigma
\Bigl[\beta\frac{1+\frac{1+\alpha\sin(\theta-\tilde\omega\tilde\tau)}{2}\Delta \tilde T}{2+\tilde\sigma(1-\cos\theta)}-\tilde{p}_{\rm air}\Bigr]\sin\theta-\tilde\Gamma\tilde\omega .\label{eq.nodim_model_omega}
\end{align}

\begin{figure*}[t!]
\centering

\setlength{\fboxsep}{0pt}

\begin{minipage}[c][0.18\textheight][c]{0.3\textwidth}
\centering
 \begin{overpic}[width=\linewidth]{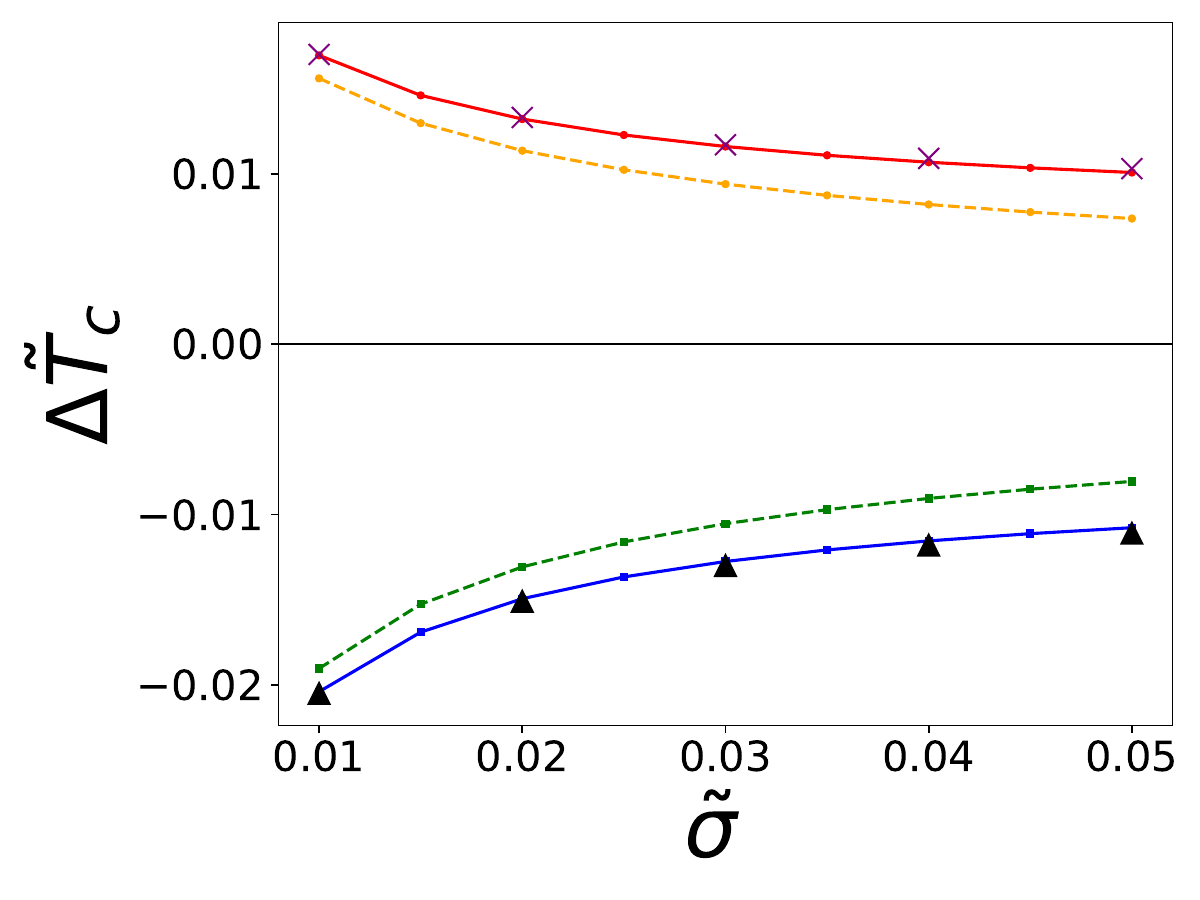}
      \put(2,70){\large (a)}
    \end{overpic}
\end{minipage}
\hfill
\begin{minipage}[c][0.18\textheight][c]{0.3\textwidth}
\centering
 \begin{overpic}[width=\linewidth]{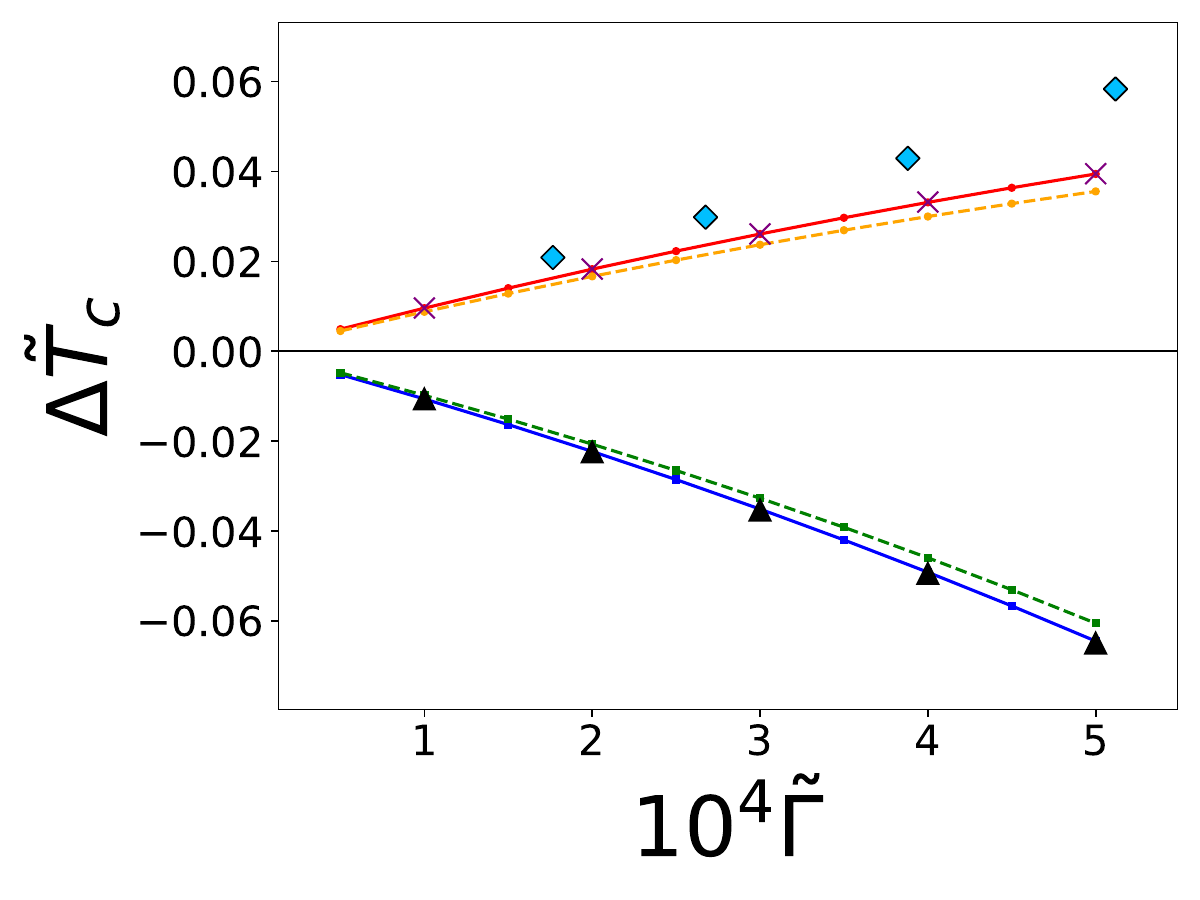}
      \put(2,70){\large (b)}
    \end{overpic}
\end{minipage}
\hfill
\begin{minipage}[c][0.18\textheight][c]{0.3\textwidth}
\centering
 \begin{overpic}[width=\linewidth]{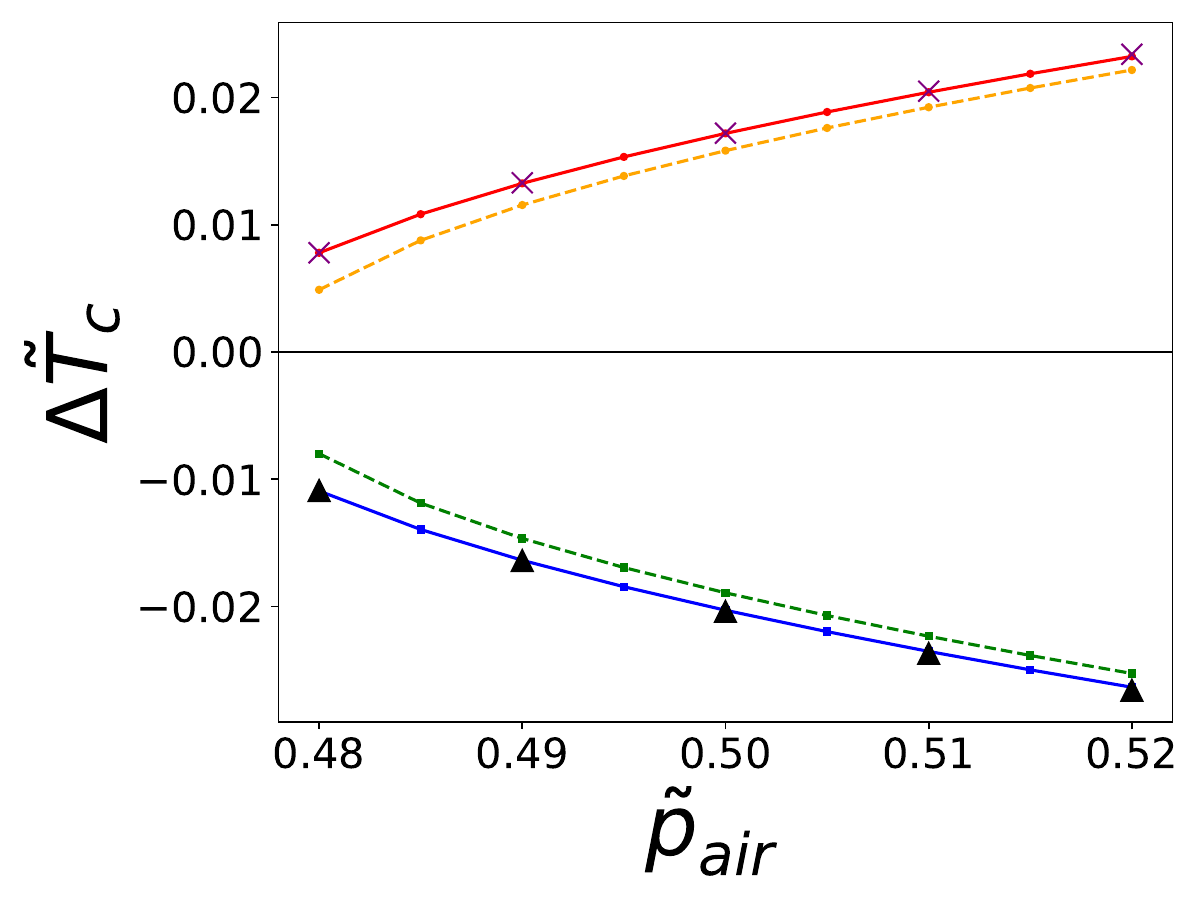}
       \put(2,70){\large (c)}
    \end{overpic}
\end{minipage}

\vspace{2mm}

\begin{minipage}[c][0.18\textheight][c]{0.3\textwidth}
\centering
 \begin{overpic}[width=\linewidth]{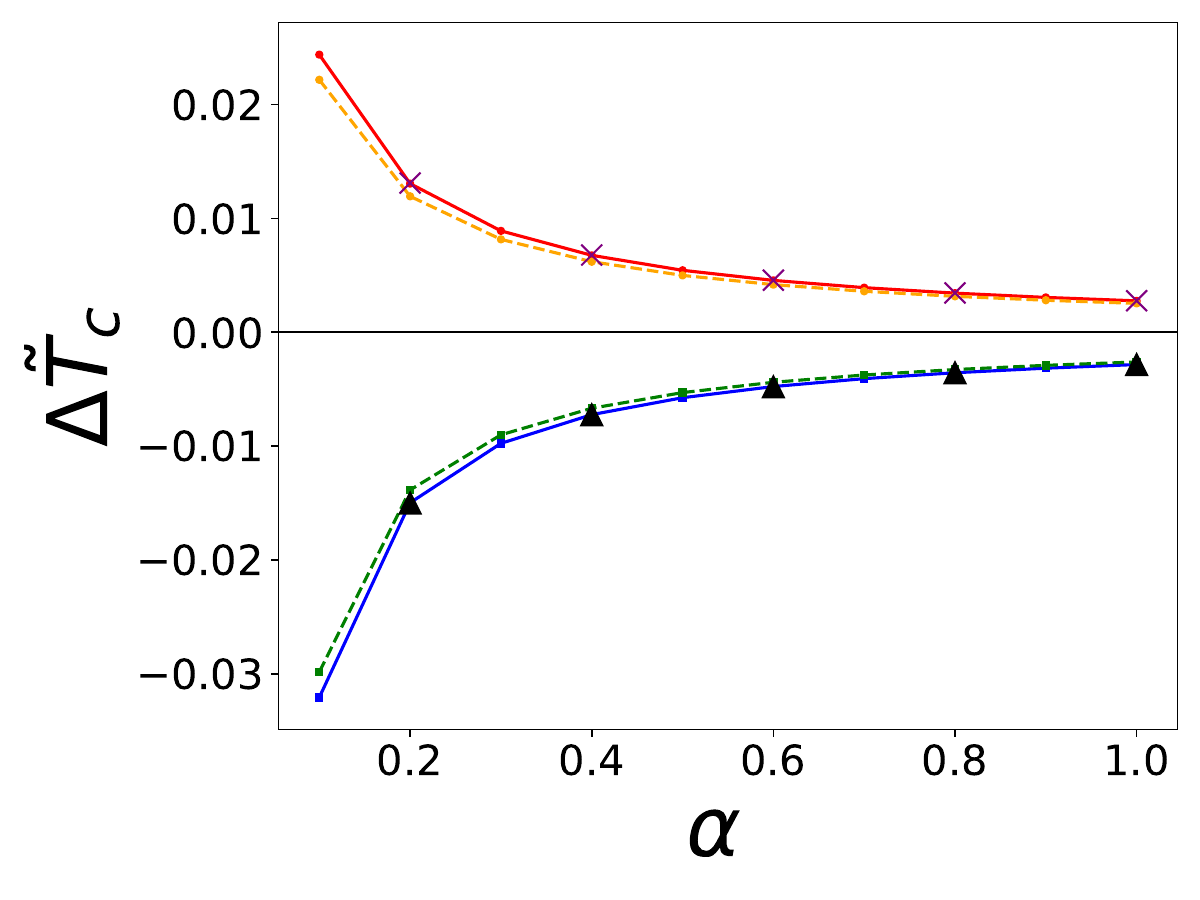}
        \put(2,70){\large (d)}
    \end{overpic}
\end{minipage}
\hfill
\begin{minipage}[c][0.18\textheight][c]{0.3\textwidth}
\centering
 \begin{overpic}[width=\linewidth]{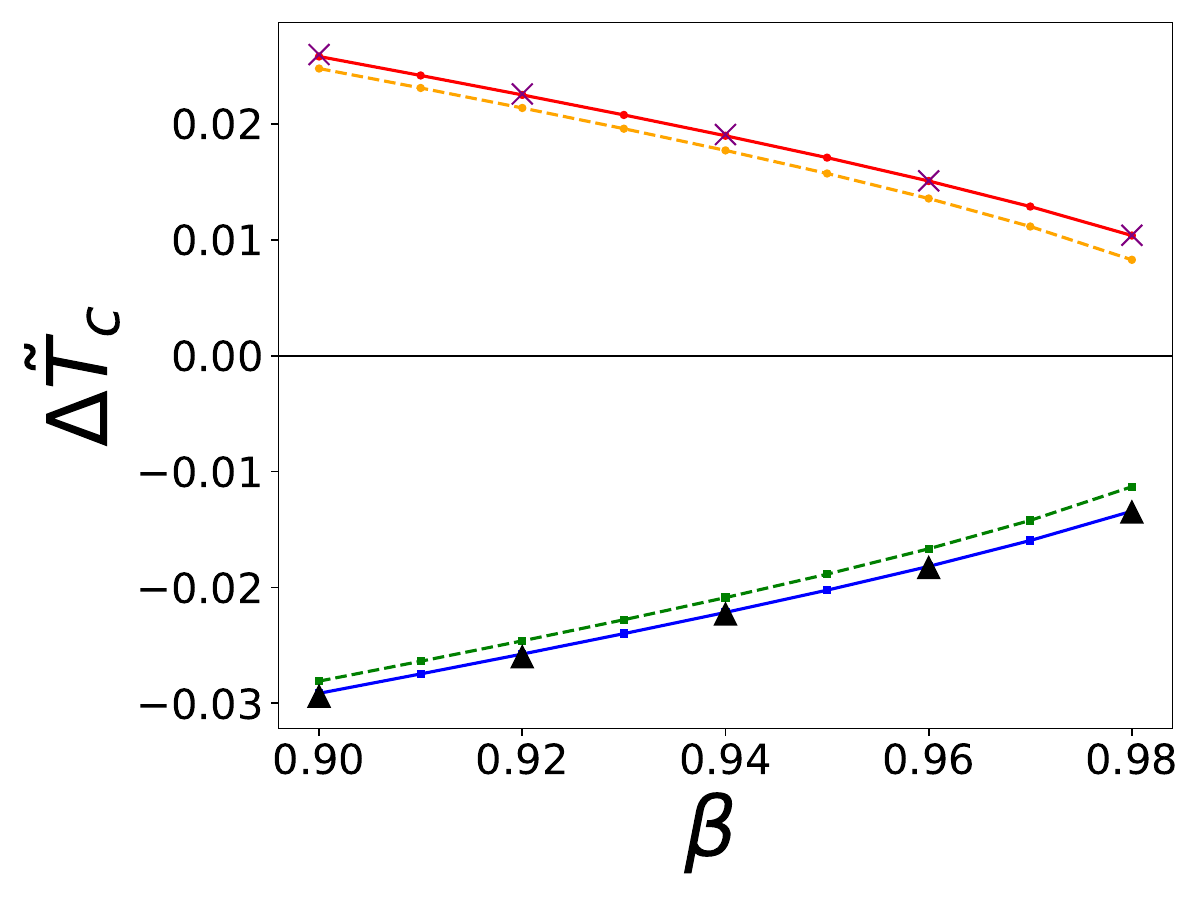}
        \put(2,70){\large (e)}
    \end{overpic}
\end{minipage}
\hfill
\begin{minipage}[c][0.18\textheight][c]{0.3\textwidth}
\centering
\includegraphics[width=0.9\linewidth]{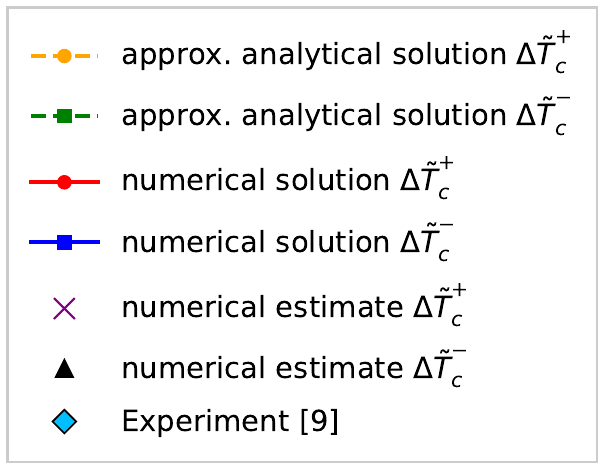}
\end{minipage}

\caption{
The critical temperature differences $\Delta \tilde T_c^\pm$ obtained by the approximate analytical solutions of Eq.~(\ref{eq.approx_solution}), numerical solutions of Eq.~(\ref{eq.numerical_solution}), and numerical estimation based on Eqs.~(\ref{eq.nodim_model_theta}) and (\ref{eq.nodim_model_omega}) plotted as functions of (a) $\tilde\sigma$, (b) $\tilde\Gamma$, (c) $\tilde p_{\rm air}$, (d) $\alpha$, and (e) $\beta$. The parameters in Eqs.~(\ref{eq.nodim_model_theta}) and (\ref{eq.nodim_model_omega}) used for the numerical calculations are given in Appendix~\ref{app:dimensionless_param}. 
The diamonds in (b) represent the experimental values of $\Delta \tilde T_c^+$ adapted from Fig.~2(d) of~\cite{toyabe2020experimental}.}
\label{fig:parameter_dependence}
\end{figure*}

The Melnikov method is applied to Eqs.~(\ref{eq.nodim_model_theta}) and (\ref{eq.nodim_model_omega}) as follows
(see Appendix~\ref{app:Melnikov} for the details).
First, the right-hand sides of Eqs.~(\ref{eq.nodim_model_theta}) and (\ref{eq.nodim_model_omega}) are decomposed into the conservative terms expressed by a Hamiltonian $H(\theta, \tilde \omega)$:
\begin{align}
H(\theta,\tilde\omega) = & \frac{\tilde\omega^2}{2}-\beta \left(1+\frac{\Delta \tilde T}{2}\right)\ln \Biggl[2 \left(1+\tilde\sigma\sin^2\frac{\theta}{2}\right)\Biggr]\nonumber\\
& -\tilde\sigma \tilde p_{\rm air}\cos\theta, \label{eq.H_0}
\end{align}
and the perturbation terms consisting of the driving and dissipation terms:
\begin{align}
g_1(\theta,\tilde\omega)=0, \quad
g_2(\theta, \tilde\omega)=\tilde \sigma \beta \frac{\frac{\alpha \sin^2 \theta}{2} \Delta \tilde T}{2+\tilde \sigma(1-\cos \theta)}-\tilde \Gamma \tilde \omega, \label{eq.perturb}
\end{align}
where we have approximated $\tilde \tau \approx 0$ because 
it has a negligible effect on $\Delta \tilde T_c^\pm$ (see Appendix~\ref{app:Melnikov} for the details).
For the conservative system, homoclinic orbits that connect $(-\pi, 0)$ and $(\pi,0)$ exist:
\begin{equation}
\begin{aligned}
\tilde\omega &=\pm\sqrt{2\beta \left(1+\frac{\Delta \tilde T}{2}\right)\ln\frac{1+\tilde\sigma\sin^2\frac{\theta}{2}}{1+\tilde\sigma}+4\tilde\sigma \tilde p_{\rm air}\cos^2\frac{\theta}{2}}\\&=\pm f(\theta),\label{eq.homoclinic_orbit_honbun}
\end{aligned}
\end{equation}
which are well-defined when $\Delta \tilde T<\frac{4\tilde\sigma \tilde p_{\rm air}}{\beta\ln{(1+\tilde\sigma)}}-2 $ is satisfied.
Here, $\tilde\omega=f(\theta)$ represents the rotational motion in which $\theta$ increases from $-\pi$ to $\pi$, whereas $\tilde\omega=-f(\theta)$ represents the rotational motion in the opposite direction, so the relevant homoclinic orbit is given by $\tilde\omega=f(\theta)$ for $\Delta \tilde T>0$ and $\tilde\omega=-f(\theta)$ for $\Delta \tilde T<0$.
When the perturbation terms are added, the homoclinic orbits of the conservative system (\ref{eq.homoclinic_orbit_honbun}) are destroyed.
However, at the homoclinic bifurcation point, the homoclinic orbit is restored.
The Melnikov function $M$ is defined as a function that represents the distance between the stable and unstable manifolds of a saddle point, which are split from the homoclinic orbit of the conservative system~\cite{Guckenheimer1983,Wiggins2003}:
\begin{align}
M
&=
\int_{-\infty}^{\infty}
\Biggl[
\frac{\partial H(\theta_0^\pm,\tilde\omega_0^\pm)}
{\partial \tilde\omega}
g_2(\theta_0^\pm,\tilde\omega_0^\pm)
\nonumber\\
&\qquad\qquad
+
\frac{\partial H(\theta_0^\pm,\tilde\omega_0^\pm)}
{\partial \theta}
g_1(\theta_0^\pm,\tilde\omega_0^\pm)
\Biggr]
\,d\tilde t ,
\label{eq.Melnikov_func}
\end{align}
where $(\theta_0^\pm(\tilde t),\tilde\omega_0^\pm(\tilde t))$ is the time-parametric representation of the homoclinic orbits (\ref{eq.homoclinic_orbit_honbun}) given by the conservative term.
When $M=0$, the stable and unstable manifolds of the saddle point approximately coincide, restoring the homoclinic orbit.
Because Eq.~(\ref{eq.Melnikov_func}) is calculated as
\begin{align}
M=\int_{\mp \pi}^{\pm \pi} \tilde \sigma \beta \frac{\frac{\alpha \sin^2 \theta}{2} \Delta \tilde T}{2+\tilde \sigma (1-\cos \theta)}d\theta
-\tilde\Gamma \int_{-\infty}^\infty {\tilde\omega_0^\pm(\tilde t)}^2 d\tilde t,
\end{align}
$M=0$ means that the driving and dissipation terms are energetically balanced.
Therefore, $\Delta \tilde T_c^\pm$ is found by solving the following equation:
\begin{align}
 \frac{(\sqrt{1+\tilde\sigma}-1)^2\alpha\beta\pi}{\tilde\sigma}\Delta \tilde T &=\pm\tilde \Gamma\int_{-\pi}^\pi f(\theta) d\theta. 
 \label{eq.numerical_solution}
\end{align}
As the right-hand side of Eq.~(\ref{eq.numerical_solution}) contains $\Delta \tilde T$, Eq.~(\ref{eq.numerical_solution}) is a self-consistent equation that determines $\Delta \tilde T$.
Since Eq.~(\ref{eq.numerical_solution}) cannot be solved analytically,
we solve it numerically or
obtain the approximate analytical solution by neglecting $O(\tilde \sigma^2)$ terms (see Sec.~\ref{app:analytical} of Appendix for the detailed derivation):
\begin{align}
\label{eq.approx_solution}
\Delta \tilde{T}_c^\pm&=\frac{32\tilde\Gamma}{\alpha^2\beta\pi^2\tilde\sigma}\Biggl[-4\tilde\Gamma\pm\sqrt{16\tilde\Gamma^2+\alpha^2\pi^2\tilde\sigma \left(\tilde{p}_{\rm air}-\frac{\beta}{2}\right)} \Biggr],
\end{align}
which is symmetric as $\Delta \tilde T_c^\pm \simeq \pm \frac{32\tilde\Gamma}{\alpha \beta\pi \sqrt{\tilde\sigma}}\sqrt{\tilde{p}_{\rm air}-\frac{\beta}{2}}$ in the linear order of $\tilde \Gamma$. Equation~(\ref{eq.approx_solution})
elucidates the basic constituents that determine the critical temperature differences and may serve as a useful and fundamental guideline for the design of the LTD Stirling engine.

\begin{table}[t!]
\caption{$\Delta \tilde T_c^\pm$ and $\Delta  T_c^\pm= \Delta \tilde T_c^\pm T_{\rm top}$ obtained using different calculation methods with the parameter values given in Appendix~\ref{app:dimensionless_param}.}
\label{tab:result}
\begin{tabular}{lllll} \hline
Equation& $\Delta \tilde T_c^+$ & $\Delta \tilde T_c^-$ & $\Delta T_c^+$& $\Delta T_c^-$\\ \hline
Eq.~(\ref{eq.approx_solution})&0.01490 &-0.01800&4.43&-5.35\\
Eq.~(\ref{eq.numerical_solution}) &0.01631&-0.01943&4.85&-5.77\\
Eqs.~(\ref{eq.nodim_model_theta}) and (\ref{eq.nodim_model_omega})&0.01636& -0.01953&4.86&-5.80\\ 
\hline
\end{tabular}
\end{table}
In Table~\ref{tab:result}, we compared the numerical solutions of Eq.~(\ref{eq.numerical_solution}), the approximate analytical solutions of Eq.~(\ref{eq.approx_solution}), and the estimation by numerically solving Eqs.~(\ref{eq.nodim_model_theta}) and (\ref{eq.nodim_model_omega}). When estimating the bifurcation points from Eqs.~(\ref{eq.nodim_model_theta}) and (\ref{eq.nodim_model_omega}), we did not ignore the $\tilde\omega\tilde\tau$ term.
As shown in Table~\ref{tab:result}, both the approximate analytical solutions and the numerical solutions are in good agreement with those obtained from the numerical estimation. In particular, the numerical solution has remarkably higher accuracy.
The asymmetry between $\Delta \tilde T_c^+$ and $\Delta \tilde T_c^-$ reflects the nonlinear effect of $\tilde \Gamma$ in Eq.~(\ref{eq.approx_solution}).

To investigate the parameter dependence, we calculated $\Delta \tilde T_c^\pm$ by varying $\tilde \sigma$, $\tilde\Gamma$, $\tilde p_{\rm air}$, $\alpha$, and $\beta$ in Fig.~\ref{fig:parameter_dependence}.
Except the varying parameter, we used the same values as those to obtain Table~\ref{tab:result}.
Similarly to Table~\ref{tab:result}, we plotted the approximate analytical solutions of Eq.~(\ref{eq.approx_solution}), the numerical solutions of Eq.~(\ref{eq.numerical_solution}), and numerical estimation using Eqs.~(\ref{eq.nodim_model_theta}) and (\ref{eq.nodim_model_omega}).
Over the entire parameter range, the values of the approximate analytical solutions and numerical solutions mostly coincide with those obtained by the numerical estimation.
Moreover, Eq.~(\ref{eq.approx_solution}) indicates that reducing $|\Delta \tilde T_c^\pm|$ requires, e.g., a smaller value of $\tilde\Gamma$, and larger values of $\tilde \sigma$ and $\alpha$.
Figure~\ref{fig:parameter_dependence} shows the same trend as this prediction. 
Furthermore, we find that $|\Delta \tilde T_c^\pm|$ decreases with $1/\alpha$ as in Fig.~\ref{fig:parameter_dependence}(d), which can be understood because $\alpha$ quantifies how effectively the given temperature difference is maintained inside the gas. Meanwhile, it may be less obvious that $|\Delta \tilde T_c^\pm|$ decreases slowly as $1/\sqrt{\tilde \sigma}$ as in Fig.~\ref{fig:parameter_dependence}(a).
We also find that the decrease in $|\Delta \tilde T_c^\pm|$ with decreasing $\tilde\Gamma$ in Fig.~\ref{fig:parameter_dependence}(b) is qualitatively consistent with the approximately linear trend of experimental data adapted from Fig.~2(d) of~\cite{toyabe2020experimental}.

\section{Outlook}
We determined the critical temperature difference below which the LTD Stirling engine ceases to rotate using a nonlinear dynamics approach.
The critical temperature difference corresponds to the homoclinic bifurcation point at which the stable limit cycle disappears. By the Melnikov method, we derived the self-consistent equations (\ref{eq.numerical_solution}) that determine the critical temperature differences and obtained both numerical and approximate analytical solutions.
In addition, from the approximate analytical solution Eq.~(\ref{eq.approx_solution}) and the parameter dependence in Fig.~\ref{fig:parameter_dependence}, we identified the physical and empirical parameters that are important for reducing the magnitude of the critical temperature differences.
This may serve as a useful and fundamental guideline for improving the LTD Stirling engine for operation under lower temperature differentials.

In principle, the present results may be verified experimentally.
For example, we can test the predicted dependence $|\Delta \tilde T_c^\pm|=O(1/\sqrt{\tilde \sigma})$.
This would require varying $\tilde\sigma=\frac{\sigma_p}{\sigma_d}$, the ratio of the sectional area of the power piston to that of the large cylinder, which is fixed for the engine reported in~[9].
One promising direction is therefore to design and build LTD Stirling engines with different values of $\tilde\sigma$, using, e.g., a 3D printer~\cite{clucas2016,Fukui2024}.
Such experiments would help test this prediction and guide the design of engines that operate at smaller temperature differences.
Furthermore, we revealed that the empirical parameters $\alpha$ and $\beta$ are key parameters governing the critical temperature differences.
However, these parameters can only be estimated by fitting to experimental data. Therefore, it is an important future direction to
develop refined dynamical models that do not rely on empirical parameters beyond the simplest model.

We expect that our work will advance the science of Stirling engines~\cite{PhysRevE.79.047702,Kitaya_2022,izumida2018nonlinear,Yin_2025,10.1093/pnasnexus/pgac251,HisashiKada2014}. 
An important remaining challenge is 
to establish a thermodynamic theory describing their efficiency and power output~\cite{10.1119/1.10023,PhysRevLett.95.190602,izumida2020quasilinear}, particularly near the critical temperature differences.
Such a theory would be valuable both for nonequilibrium physics and for the practical design of LTD Stirling engines.

\begin{acknowledgments}
The authors used ChatGPT (OpenAI, GPT-5.5, GPT-5.6, and GPT-6.1) to assist with manuscript editing, consistency check, and developing numerical codes used for the numerical calculations.
\end{acknowledgments}

% -----------------------------
% References
% -----------------------------
\bibliographystyle{apsrev4-2}

\bibliography{references}

\onecolumngrid

\appendix
\section{Details of numerical calculations}
\label{app:plot_value}
We used the following physical parameters to obtain the results in Fig.~\ref{fig:experiment_model}~\cite{toyabe2020experimental}:
$2r\sigma_d=\SI{44900}{mm^3}$, $\sigma_p=\SI{71}{mm^2}$, $r=\SI{3.5}{mm}$, $I=5.7\times10^{-5}\si{kgm^2}$, $p_{\rm air}=\SI{101.3}{kPa}$, $n=\SI{0.00185}{mol}$, $R=\SI{8.314}{J/Kmol}$, $T_{\rm top}=297.15\si{K}$, and $\Gamma = 0.05\,I\,\si{\per\second}$. 
Also, the empirical parameters $\alpha=0.157$, $\beta=0.954$, and $\tau=\SI{13.7}{ms}$ were used, 
estimated from the experimental data~\cite{toyabe2020experimental}, using maximum likelihood estimation.
We used the Tsit5 solver in the Julia package DifferentialEquations.jl for the numerical calculations of the differential equations, except those in Fig.~\ref{fig:phase_diagram}, which were performed using the DOP853 solver in the Python package Scipy. We also used brentq function, which implements Brent method in the Python package SciPy, to calculate the numerical solutions of Eq.~(\ref{eq.numerical_solution}).

\section{Derivation of self-consistent equation~(\ref{eq.numerical_solution}) using Melnikov method}
\label{app:Melnikov}
In this Section, we derive the self-consistent equations (Eq.~(\ref{eq.numerical_solution}) in the main text) that determine the critical temperature differences. The target equations are as follows (Eqs.~(\ref{eq.nodim_model_theta}) and (\ref{eq.nodim_model_omega}) in the main text):
\begin{align}
\dot{\theta} &= \tilde\omega,\label{eq.theta_omegatau}\\
\dot{\tilde{\omega}}&=
\tilde\sigma\Biggl[\beta\frac{1+\frac{1+\alpha\sin(\theta-\tilde\omega\tilde\tau)}{2}\Delta \tilde T}{2+\tilde\sigma(1-\cos\theta)}-\tilde{p}_{\rm air}\Biggr]\sin\theta-\tilde\Gamma\tilde\omega.\label{eq.omega_omegatau}
\end{align}
As we will justify later, even if we ignore the effect of $\tilde \tau$, it is not influential on the bifurcations points. Therefore, we focus on the following equations by putting $\tilde \tau=0$ into Eqs.~(\ref{eq.theta_omegatau}) and (\ref{eq.omega_omegatau}):
\begin{align}
\dot{\theta} &= \tilde\omega,\label{eq.theta}\\
\dot{\tilde{\omega}}&=
\tilde\sigma\Biggl[\beta\frac{1+\frac{1+\alpha\sin\theta}{2}\Delta \tilde T}{2+\tilde\sigma(1-\cos\theta)}-\tilde{p}_{\rm air}\Biggr]\sin\theta-\tilde\Gamma\tilde\omega.\label{eq.omega}
\end{align}
We apply the Melnikov method to Eqs.~(\ref{eq.theta}) and (\ref{eq.omega}), which is a standard method for analytically determining approximate homoclinic bifurcation points~\cite{Guckenheimer1983,Wiggins2003}.
Here, we focus on the homoclinic bifurcation 
where a part of a limit cycle collides with the saddle point at $(\pm \pi,0)$ and disappears.
Here, $(\pi,0)$ and $(-\pi,0)$ are identified as the same point due to $2\pi$-periodicity.

The right-hand sides of Eqs.~(\ref{eq.theta}) and (\ref{eq.omega}) are decomposed into the conservative terms expressed by a Hamiltonian $H$, and the perturbation terms consisting of the driving and dissipation terms:
\begin{align}
&\dot \theta=\frac{\partial H(\theta, \tilde\omega)}{\partial \tilde \omega}+g_1(\theta,\tilde\omega),\label{eq.theta_g1}\\
&\dot{\tilde{\omega}}=-\frac{\partial H(\theta, \tilde\omega)}{\partial \theta}+g_2(\theta,\tilde\omega).\label{eq.omega_g2}
\end{align}
Here, the Hamiltonian of the conservative system is given by
\begin{align}
H(\theta,\tilde\omega) &=\frac{\tilde\omega^2}{2}-\beta \left(1+\frac{\Delta \tilde T}{2}\right)\ln \Biggl[2 \left(1+\tilde\sigma\sin^2\frac{\theta}{2}\right)\Biggr]-\tilde\sigma \tilde p_{\rm air}\cos\theta, \label{eq.H_0}
\end{align}
and the perturbation terms are given by
\begin{align}
g_1(\theta,\tilde\omega)=0, \quad
g_2(\theta, \tilde\omega)=\tilde \sigma \beta \frac{\frac{\alpha \sin^2 \theta}{2} \Delta \tilde T}{2+\tilde \sigma(1-\cos \theta)}-\tilde \Gamma \tilde \omega, \label{eq.perturb}
\end{align}
where $g_2(\theta, \tilde \omega)$ consists of the driving term that is proportional to $\Delta \tilde T$ and the dissipation term $-\tilde \Gamma \tilde \omega$ due to friction.
Here, it should be noted that $\Delta \tilde T$ is included in both the Hamiltonian of the conservative system Eq.~(\ref{eq.H_0}) and the perturbation terms Eq.~(\ref{eq.perturb}).

We first consider the conservative system governed by the Hamiltonian $H$. In that system, the homoclinic orbits that connect the saddle points $(-\pi,0)$ and $(\pi, 0)$ exist.
Using 
$H(-\pi,0)=H(\pi,0)=-\beta \left(1+\frac{\Delta \tilde T}{2}\right)\ln[2(1+\tilde \sigma)]+\tilde \sigma \tilde p_{\rm air}$, we obtain
\begin{align}
 \frac{\tilde\omega^2}{2}-\beta \left(1+\frac{\Delta \tilde T}{2}\right)\ln \Biggl[2\left(1+\tilde\sigma\sin^2\frac{\theta}{2}\right)\Biggr]-\tilde\sigma \tilde p_{\rm air}\cos\theta=-\beta \left(1+\frac{\Delta \tilde T}{2}\right)\ln[2(1+\tilde\sigma)]+\tilde\sigma\tilde p_{\rm air}. 
 \label{eq.derivation_homoclinic_orbit}
\end{align}
Solving Eq.~(\ref{eq.derivation_homoclinic_orbit}) for $\tilde\omega$, we derive the homoclinic orbits on the phase plane:
\begin{align}
\tilde\omega &=\pm\sqrt{2\beta \left(1+\frac{\Delta \tilde T}{2}\right)\ln\frac{1+\tilde\sigma\sin^2\frac{\theta}{2}}{1+\tilde\sigma}+4\tilde\sigma \tilde p_{\rm air}\cos^2\frac{\theta}{2}}=\pm f(\theta).\label{eq.homoclinic_orbit}
\end{align}
Because the expression under the square root must be positive, these orbits are well-defined when $\Delta \tilde T < \frac{4\tilde\sigma \tilde p_{\rm air}}{\beta\ln{(1+\tilde\sigma)}}-2 $ is satisfied.
Since $\tilde\omega=\pm f(\theta)$ means the phase change from $\mp\pi$ to $\pm\pi$, the relevant homoclinic orbit is given by $\tilde\omega=f(\theta)$ for $\Delta \tilde T>0$ and $\tilde\omega=-f(\theta)$ for $\Delta \tilde T<0$.

Next, we consider the case where the perturbation terms are added. In general, while the homoclinic orbit of the conservative system is destroyed with the addition of the perturbation terms, the homoclinic orbit may be restored at the bifurcation point.
In the Melnikov method, this bifurcation point is approximately determined at which the following Melnikov function $M$ vanishes~\cite{Guckenheimer1983,Wiggins2003}:
\begin{align}
M=\int_{-\infty}^\infty \left(\frac{\partial H(\theta, \tilde \omega)}{\partial \tilde \omega}g_2(\theta, \tilde\omega)+\frac{\partial H(\theta, \tilde\omega)}{\partial \theta}g_1(\theta, \tilde\omega)\right)\Biggl |_{(\theta(\tilde t),\tilde\omega(\tilde t))=(\theta_0^{\pm}(\tilde t),\tilde\omega_0^{\pm}(\tilde t))} d\tilde t,\label{eq.Melnikov_func_app}
\end{align}
where $(\theta(\tilde t),\tilde\omega(\tilde t))=(\theta_0^{\pm}(\tilde t),\tilde\omega_0^{\pm}(\tilde t))$ is the time-parametric representation of the homoclinic orbit Eq.~(\ref{eq.homoclinic_orbit}).
The Melnikov function describes the distance between the stable and unstable manifolds of the saddle point $(\pm \pi, 0)$, which are separated because the homoclinic orbit of the original conservative system is destroyed due to the perturbation.
When Eq.~(\ref{eq.perturb}) is applied to Eq.~(\ref{eq.Melnikov_func_app}), we have
%Eq.~(\ref{eq.Melnikov_func})にEq.~(\ref{eq.perturb})を用いると、メルニコフ関数は
\begin{align}
M&=\int_{-\infty}^\infty \tilde\omega g_2(\theta, \tilde\omega)|_{(\theta(\tilde t),\tilde\omega(\tilde t))=(\theta_0^{\pm}(\tilde t),\tilde\omega_0^{\pm}(\tilde t))}d\tilde t\nonumber\\
&=\int_{-\infty}^\infty \left(\tilde \sigma \beta \frac{\frac{\alpha \sin^2 \theta}{2} \Delta \tilde T}{2+\tilde \sigma (1-\cos \theta)} \tilde\omega-\tilde\Gamma \tilde\omega^2\right)\Biggl |_{(\theta(\tilde t),\tilde\omega(\tilde t))=(\theta_0^{\pm}(\tilde t),\tilde\omega_0^{\pm}(\tilde t))} d\tilde t \nonumber\\
&=\int_{\mp \pi}^{\pm \pi} \tilde \sigma \beta \frac{\frac{\alpha \sin^2 \theta}{2} \Delta \tilde T}{2+\tilde \sigma (1-\cos \theta)}d\theta-\tilde\Gamma \int_{-\infty}^\infty {\tilde\omega_0^\pm(\tilde t)}^2 d\tilde t \nonumber\\
&=\pm\frac{(\sqrt{1+\tilde\sigma}-1)^2\alpha\beta\pi}{\tilde\sigma}\Delta \tilde T-\tilde\Gamma \int_{-\infty}^\infty {\tilde\omega_0^\pm(\tilde t)}^2 d\tilde t.
\end{align}
Therefore, the condition $M=0$ represents the energy balance between the driving and dissipative terms.
By using Eq.~(\ref{eq.homoclinic_orbit}) and $\tilde \omega=d\theta/d\tilde t$, we have
\begin{align}
\int_{-\infty}^\infty \tilde \omega_0^\pm(\tilde t)^2 d\tilde t&=\int_{-\pi}^{\pi} f(\theta) d\theta.
\end{align}
Thus, the self-consistent equations determining the homoclinic bifurcation points are given as follows:
\begin{align}
\frac{(\sqrt{1+\tilde\sigma}-1)^2\alpha\beta\pi}{\tilde\sigma}\Delta \tilde T &= \pm\tilde \Gamma\int_{-\pi}^{\pi} f(\theta) d\theta, \label{eq.M=0}
\end{align}
which is Eq.~(\ref{eq.numerical_solution}) in the main text.

Subsequently, we justify neglecting effect of $\tilde\tau$ in Eqs.~(\ref{eq.theta_omegatau}) and (\ref{eq.omega_omegatau}). Instead of Eqs.~(\ref{eq.theta}) and (\ref{eq.omega}), we consider Eqs.~(\ref{eq.theta_omegatau}) and (\ref{eq.omega_omegatau}).
As the Hamiltonian of the conservative system is the same as Eq.~(\ref{eq.H_0}), the corresponding homoclinic orbit of the conservative system is equal to Eq.~(\ref{eq.homoclinic_orbit}).
The Melnikov function $M'$ is then given by 
\begin{align}
M'&=\int_{\mp \pi}^{\pm \pi} \tilde \sigma \beta \frac{\frac{\alpha \sin \theta\sin(\theta-\tilde\omega\tilde\tau)}{2} \Delta \tilde T}{2+\tilde \sigma (1-\cos \theta)}d\theta-\tilde\Gamma \int_{-\infty}^\infty {\tilde\omega_0^\pm(\tilde t)}^2 d\tilde t.
\end{align}
Because we may make a first-order approximation $\sin(\theta-
\tilde\omega\tilde\tau)\simeq\sin\theta-\tilde\omega\tilde\tau\cos\theta$ 
for $\tilde\omega\tilde\tau$
for the parameter values we have studied, we have
\begin{align}
\int_{\mp \pi}^{\pm \pi} \tilde \sigma \beta \frac{\frac{\alpha \sin\theta\sin(\theta-\tilde\omega\tilde\tau)}{2} \Delta \tilde T}{2+\tilde \sigma (1-\cos \theta)}d\theta &\simeq \int_{\mp \pi}^{\pm \pi} \tilde \sigma \beta \frac{\frac{\alpha \sin^2 \theta}{2} \Delta \tilde T}{2+\tilde \sigma(1-\cos \theta)}d\theta
-\underbrace{\int_{\mp \pi}^{\pm \pi} \tilde \sigma \beta \frac{\frac{\alpha \tilde \omega \tilde\tau \sin \theta \cos \theta}{2} \Delta \tilde T}{2+\tilde \sigma (1-\cos \theta)}d\theta}_{=0}
\nonumber\\
&=\pm\frac{(\sqrt{1+\tilde\sigma}-1)^2\alpha\beta\pi}{\tilde\sigma} \Delta \tilde T,
\end{align}
which implies
\begin{align}
M'&=\pm\frac{(\sqrt{1+\tilde\sigma}-1)^2\alpha\beta\pi}{\tilde\sigma}\Delta \tilde T-\tilde\Gamma \int_{-\infty}^\infty {\tilde\omega_0^\pm(\tilde t)}^2 d\tilde t = M. 
\end{align}
Thus, the bifurcation points obtained with $\tilde\tau=0$ coincide with those obtained by considering $\tilde\tau$ to the first order, justifying the neglect of the effect of $\tilde \tau$.
In fact, for the parameter values given in Appendix~D, when we compared two values, one of which was calculated from Eqs.~(\ref{eq.theta_omegatau}) and (\ref{eq.omega_omegatau}) considering $\tilde\tau$ and the other was calculated from the same equations with $\tilde\tau=0$ (i.e., Eqs.~(\ref{eq.theta}) and (\ref{eq.omega})), their relative error was less than 1$\%$, which supports our justification.

\section{Derivation of approximate analytical solution (\ref{eq.approx_solution})}\label{app:analytical}
While obtaining exact analytical solutions of Eq.~(\ref{eq.M=0}) (Eq.~(\ref{eq.numerical_solution}) in the main text) may be difficult, we may obtain its approximate analytical solutions by neglecting $O(\tilde \sigma^2)$ terms of Eq.~(\ref{eq.M=0}).
In fact, as a compression ratio of an LTD Stirling engine is very small~\cite{Senft2000}, in other words $\tilde \sigma \ll 1$, 
this approximation is considered to be reasonable.
Since $\sqrt{1+x}\simeq1+\frac{x}{2}$ for $|x| \ll 1$, the left-hand side of Eq.~(\ref{eq.M=0}) is approximated as 
\begin{align}
\frac{(\sqrt{1+\tilde\sigma}-1)^2\alpha\beta\pi}{\tilde\sigma}\Delta \tilde T&\simeq  \frac{\alpha\beta\pi\tilde \sigma}{4} \Delta \tilde T.\label{eq.left_sec}
\end{align}
Next, we consider the right-hand side $\pm \tilde \Gamma\int_{-\pi}^\pi f(\theta) d\theta$.
We apply the first-order approximation of $\tilde\sigma$ to 
$f(\theta)=\sqrt{2\beta \left(1+\frac{\Delta \tilde T}{2}\right)\ln\frac{1+\tilde\sigma\sin^2\frac{\theta}{2}}{1+\tilde\sigma}+4\tilde\sigma \tilde p_{\rm air}\cos^2\frac{\theta}{2}}$. 
Using $\frac{1}{1+x}\simeq1-x$, we have
\begin{align}
\frac{1+\tilde\sigma\sin^2\frac{\theta}{2}}{1+\tilde\sigma} &\simeq \left(1+\tilde \sigma\sin^2\frac{\theta}{2}\right)(1-\tilde \sigma)\nonumber\\&\simeq1+\tilde \sigma\sin^2\frac{\theta}{2}-\tilde \sigma \nonumber\\&=1-\tilde \sigma\cos^2\frac{\theta}{2}.
\end{align}
Then, using $\ln(1-x)\simeq-x$, we also have
%より
\begin{align}
    \ln \frac{1+\tilde\sigma\sin^2\frac{\theta}{2}}{1+\tilde\sigma} \simeq-\tilde\sigma\cos^2\frac{\theta}{2}.
\end{align}
Thus, we obtain
% よって、
\begin{align}
    f(\theta)&\simeq\sqrt{2\beta \left(1+\frac{\Delta \tilde T}{2}\right)\left(-\tilde\sigma\cos^2\frac{\theta}{2}\right)+4\tilde\sigma\tilde p_{\rm air}\cos^2\frac{\theta}{2}}\nonumber\\& =
    2\sqrt{\tilde\sigma \left(\tilde p_{\rm air}-\beta\frac{1+ \frac{\Delta \tilde T}{2}}{2}\right)}\cos\frac{\theta}{2}.\label{eq.right_sec}
\end{align}
By using Eqs.~(\ref{eq.left_sec}) and (\ref{eq.right_sec}), the self-consistent equations Eq.~(\ref{eq.M=0}) are reduced to
the following equations:
\begin{align}
\frac{\alpha\beta\tilde\sigma\pi\Delta \tilde T}{4} &=8\tilde\Gamma\sqrt{\tilde\sigma \left(\tilde p_{\rm air}-\beta\frac{1+\frac{\Delta \tilde T}{2}}{2}\right)},\label{eq.sec_+_approx}
\end{align}
\begin{align}
\frac{\alpha\beta\tilde\sigma\pi\Delta \tilde T}{4} &=-8\tilde \Gamma\sqrt{\tilde\sigma \left(\tilde p_{\rm air}-\beta\frac{1+\frac{\Delta \tilde T}{2}}{2}\right)}.\label{eq.sec_-_approx}
\end{align}
Squaring these equations reproduce the same quadratic equation with respect to $\Delta \tilde T$. The positive solution ($\Delta \tilde T>0$) is the solution of Eq.~(\ref{eq.sec_+_approx}), and the negative solution ($\Delta \tilde T<0$) is the solution of Eq.~(\ref{eq.sec_-_approx}).
Denoting the solutions of Eqs.~(\ref{eq.sec_+_approx}) and (\ref{eq.sec_-_approx}) by $\Delta \tilde{T}_c^+$ and $\Delta \tilde{T}_c^-$, respectively, we finally obtain
\begin{align}
\Delta \tilde{T}_c^\pm&=\frac{32\tilde\Gamma}{\alpha^2\beta\pi^2\tilde\sigma}\Biggl[-4\tilde\Gamma \pm \sqrt{16\tilde\Gamma^2+\alpha^2\pi^2\tilde\sigma \left(\tilde{p}_{\rm air}-\frac{\beta}{2}\right)} \Biggr],
\end{align}
which is Eq.~(14) in the main text.

\section{Nondimensionalized quantities}
\label{app:dimensionless_param}
By nondimensionalizing the physical parameters in Appendix~\ref{app:plot_value}, we obtain the following nondimensionalized parameters in Eqs.~(\ref{eq.nodim_model_theta}) and (\ref{eq.nodim_model_omega}):
$\tilde\Gamma\simeq0.0001766$, $\tilde\sigma\simeq0.01107$, $\tilde p_{\rm air}\simeq0.4976$, and $\tilde\tau\simeq3.879$.
As for the empirical parameters $\alpha$ and $\beta$, we used the same values as in Appendix~\ref{app:plot_value} as they are originally dimensionless.

\end{document}